\documentclass[aps,prb,twocolumn,amsmath,amssymb,superscriptaddress]{revtex4-1}

\usepackage{multirow}%
\usepackage{graphicx}
\usepackage{dcolumn}
\usepackage{bm}
\usepackage[varg]{txfonts}
\usepackage{natbib}%
\newcommand{\lco}{La$_2$CuO$_4$ }

\newcommand{\scoc}{Sr$_2$CuO$_2$Cl$_2$ }
\newcommand{\nco}{Nd$_2$CuO$_4$ }

\renewcommand{\vec}[1]{\mbox{\boldmath$#1$}}

\begin{document}


\title{Polarization dependence and symmetry analysis in indirect $K$-edge RIXS}

\author{G. Chabot-Couture}
\affiliation{Department of Applied Physics, Stanford University, Stanford, California 94305, USA}
\altaffiliation[Corresponding author: ]{gchabcou@stanford.edu}

\author{J. N. Hancock}
\affiliation{D\'epartement de Physique de la Mati\`ere Condens\'ee, Universit\'e de Gen\`eve, Gen\`eve, CH-1211, Switzerland}

\author{P. K. Mang}
\affiliation{Department of Applied Physics, Stanford University, Stanford, California 94305, USA}

\author{D. M. Casa}
\affiliation{Advanced Photon Source, Argonne National Laboratory, Argonne, Illinois 60439, USA}

\author{T. Gog}
\affiliation{Advanced Photon Source, Argonne National Laboratory, Argonne, Illinois 60439, USA}

\author{M. Greven}
\affiliation{School of Physics and Astronomy, University of Minnesota, Minneapolis, Minnesota 55455, USA}

\date{\today}

\begin{abstract}
We present a study of the charge-transfer excitations in undoped Nd$_2$CuO$_4$ using resonant inelastic X-ray scattering (RIXS) at the Cu $K$-edge. At the Brillouin zone center, azimuthal scans that rotate the incident-photon polarization within the CuO$_2$ planes reveal weak fourfold oscillations. A comparison of spectra taken in different Brillouin zones reveals a spectral weight decrease at high energy loss from forward- to back-scattering. We show that these are scattered-photon polarization effects related to the properties of the observed electronic excitations. Each of the two effects constitutes about 10\% of the inelastic signal while the `$4p$-as-spectator' approximation describes the remaining 80\%. Raman selection rules can accurately model our data, and we conclude that the observed polarization-dependent RIXS features correspond to $E_g$ and $B_{1g}$ charge-transfer excitations to non-bonding oxygen $2p$ bands, above $2.5$~eV energy-loss, and to an $E_g$ $d\rightarrow d$ excitation at $1.65$~eV.
\end{abstract}

\date{\today}

\maketitle

Raman scattering and optical spectroscopy have enabled tremendous contributions to the study of condensed matter systems. Both probes use $\sim1$~eV light and are limited to essentially zero momentum transfer. In order to investigate the charge response of a material throughout the Brillouin zone, photons in the X-ray regime must be used. X-ray Raman scattering, more commonly referred to as resonant inelastic X-ray scattering (RIXS), allows the measurement of the momentum dependence of charge excitations. Even though it has successfully been used to study the physics of a wide array of systems,\cite{greniers:2005, hilljp:2008, huotaris:2008, lul:2005, hancockjn:2009} this resonant technique is relatively new and there is still much to learn about the details of its cross section.

One of the strengths of conventional Raman scattering derives from the fact that the technique allows the determination of the symmetry of excitations by selecting the polarization of the incident and scattered photons. Only polarization dependent soft (or direct~\cite{vandenbrinkj:2006}) X-ray RIXS studies, at the O $K$-edge or the Cu $L$- and $M$-edges of the cuprates,\cite{okadak:2002, haraday:2002, dudalc:1998, ghiringhellig:2004, kuiperp:1998} have shown that photon polarization can be used to select different electronic excitations. In these cases, the photon polarization effects are understood to come from the direct interaction of the photoelectron, excited into the valence system, with the valence electrons. This is why this type of RIXS is commonly referred to as direct.\cite{vandenbrinkj:2006} On the other hand, tuning the incident energy to the Cu $K$-edge excites the photoelectron into the Cu $4p$ band approximately $10$-$20$~eV above the $3d$ valence levels. The relatively large separation in energy from the valence levels as well as the large spatial extent of the $4p$ orbital are believed to prevent the photoelectron from interacting with the valence system, and this type of RIXS is in turn referred to as indirect.\cite{vandenbrinkj:2006} Accordingly, theoretical models of this type of RIXS accordingly do not include possible interactions of the $4p$ photoelectron with the valence system,\cite{tsutsuik:1999, igarashij:2006, vernayf:2008, amentljp:2007} which is commonly referred to as the `$4p$-as-spectator' approximation.

RIXS investigations of the photon polarization effects have so far concluded that the incident polarization does not affect the probed valence band excitations, but only determines their specific resonance energies based on the crystal field levels of the $4p$ photoelectron,\cite{hamalainenk:2000, kimyj:2007, vernayf:2008} in accordance with the $4p$-as-spectator approximation. Nonetheless, there still exists no quantitative experimental evidence for an incident-photon polarization independence. Furthermore, the scattered-photon polarization effects at the transition metal $K$-edge have received very little attention, and it remains unclear whether they are present and can be used to learn about the symmetry of charge-transfer excitations in transition metal oxides.

This paper is separated in three parts. In Section \ref{sc:pol}, we discuss the photon-polarization dependence expected within the $4p$-as-spectator approximation. In Sec. \ref{sc:nco}, we study the photon-polarization and scattering-geometry dependence of RIXS at the Cu $K$-edge of Nd$_2$CuO$_4$ and observe scattered-photon polarization effects beyond the $4p$-as-spectator approximation. In Sec. \ref{ssc:nco:psi}  we present data obtained upon rotating the incident-photon polarization within the CuO$_2$ planes while in Sec. \ref{ssc:nco:scatgeom} we present a comparison of zone-center spectra taken in different Brillouin zones. The results are discussed in Sec. \ref{sc:disc}. The normalization procedure used to correct for sample self-absorption and compare RIXS signal across different scattering geometries is described in Appendix \ref{sc:norm}.

\section{Polarization dependence of the RIXS cross section\label{sc:pol}}

Before presenting our experimental observations of photon polarization effects in RIXS, it is instructive to discuss what is expected according the current understanding of the indirect RIXS cross section. At the Cu $K$-edge, the RIXS process starts with the resonant absorption of an X-ray photon which creates a $1s$ core hole and a $4p$ photoelectron on the Cu site. During the lifetime of the core hole, before it recombines with the photoelectron and an X-ray photon is emitted, the core-hole interacts strongly with the valence system to create electronic excitations. The $4p$ photoelectron, on the other hand, is believed to be only a spectator and to evolve without interacting with the valence system.

In order to determine the cross-sectional implications of this $4p$-as-spectator approximation, we consider a simple model where the photoelectron's energy eigenstates are dictated by the crystal-field: $E_{x,y}^{4p}=\Delta$ and $E_z^{4p}=0$. While this approach neglects $4p$ band effects, in practice, the RIXS intensity is often confined within narrow and well-separated intervals in incident energy which should be approximated well by this two-level model. Since the initial and final states of the $4p$ photoelectron are determined by the initial ($\vec{\epsilon}_i$) and scattered ($\vec{\epsilon}_f$) photon polarizations, the transition amplitude ($I_{4p}$) of the photoelectron is determined by its evolution within the crystal-field during the lifetime of the core hole ($\hbar/\Gamma$) and contains polarization dependent factors that modulate the intensity of the excitations created by the core-hole:
\begin{align}
I_{4p}&=\left|\left\langle\vec{\epsilon}_f\right|\frac{1}{E_i-H-i\Gamma}\left|\vec{\epsilon}_i\right\rangle\right|^2\nonumber\\
&=\left|\frac{\epsilon_f^x\epsilon_i^x+\epsilon_f^y\epsilon_i^y}{E_i-\Delta-i\Gamma}+\frac{\epsilon_f^z\epsilon_i^z}{E_i-i\Gamma}\right|^2\label{eqn:i4p}
\end{align}
In the case where $\Delta\gg\Gamma$, the cross section simplifies to a polarizer: only the polarization components along the resonating $4p$ crystal-field level contribute to the inelastic signal: $\left|\epsilon_f^x\epsilon_i^x+\epsilon_f^y\epsilon_i^y\right|^2$ for an in-plane resonance ($E_i=\Delta$) and $|\epsilon_f^z\epsilon_i^z|^2$ for an out-of-plane resonance ($E_i=0$).

In their study of CuO, \citet{doringg:2004} pointed out that, within the $4p$-as-spectator approximation, the electric dipole absorption-emission matrix element is equivalent to the resonant elastic X-ray scattering cross section described in detail by \citet{hannonjp:1988}. Furthermore, they successfully apply this approximation to describe the scattering angle dependence of the intensity of the 5.4~eV local charge-transfer excitation in horizontal scattering geometry. Equation~\ref{eqn:i4p} is a simplification of the formula presented in Ref.~\citep{hannonjp:1988} but it succinctly captures the important effect of scattering geometry on the RIXS signal. Including different valence states, like the well- and poorly-screened intermediate states, would improve it.

In general, Eq.~\ref{eqn:i4p} shows that the inelastic intensity is maximized when both the incident- and scattered-photon polarizations are parallel to each other and point along a crystal-field eigenstate. These conditions can naturally be fulfilled in vertical-scattering geometry where both polarizations are perpendicular to the scattering plane ($\sigma$ polarized).\cite{hilljp:2008,kimyj:2007} For horizontal scattering geometry, with both polarizations within the scattering plane ($\pi$ polarized), these conditions can only be approached for forward and backward scattering.\cite{collarte:2006,hasanmz:2002}

On the other hand, in horizontal-scattering geometry the elastic intensity can be minimized independently of the inelastic intensity. This can be an advantage since it allows the suppression of the elastic `tail' due to the non-zero energy resolution, leading to a higher signal-to-background ratio. At a scattering angle ($2\theta$) of 90$^\circ$, the non-resonant elastic contribution is zero because the polarization factor of Thompson scattering ($\vec{\epsilon}_f\cdot\vec{\epsilon}_i$) is zero. To have non-zero inelastic intensity, the non-degenerate $4p$ eigenstates can be used as cross-polarizers. By polarizing the incident photon between $4p$ crystal-field eigenstates of different energies, e.g., $\vec{\epsilon}_i\:||\:\vec{x}+\vec{z}$, the inelastic signal will be dominated by the excitations created by the incident photon polarization component along the resonating $4p$ crystal-field eigenstate. Since this resonating component is not perpendicular to the scattered-photon polarization, the resonant inelastic signal will be detectable while the non-resonant elastic signal will be zero. Note that this effectively rotates the photon polarization 90$^\circ$. This should not be mistaken for a scattered-photon polarization effect where the $4p$ photoelectron scatters during the RIXS process.

This type of scattering geometry is used extensively.\cite{ishiik:2005,lul:2006} Based solely on the $4p$ polarization effects, the maximum RIXS intensity in this horizontal scattering geometry is observed when both incident and scattered photon polarization are at $45^{\circ}$ from the resonantly excited $4p$ eigenstate. In the tetragonal crystal-field symmetry of cuprates for example, $\vec{Q}\:||\:\vec{c}$ and $2\theta=90^\circ$ allows maximum RIXS intensity for both in-plane and out-of-plane resonances while minimizing the non-resonant contribution to the elastic line. As is shown in Appendix \ref{sc:norm}, self-absorption effects will move this maximum of RIXS intensity towards a more grazing incidence angle (keeping $2\theta=90^\circ$) when the scattering surface is perpendicular to the $[0~0~L]$ direction.

\section{Photon polarization effects in NCO\label{sc:nco}}

For our study of the scattering-geometry and photon-polarization dependence of RIXS, we chose Nd$_2$CuO$_4$ (NCO), the tetragonal Mott-insulating parent compound of the electron-doped high-temperature superconductor Nd$_{2-x}$Ce$_{x}$CuO$_4$. A single crystal was prepared as described previously~\cite{mangpk:2004} and studied in its as-grown state. A larger piece from the same growth was measured with neutron scattering and found to have a N\'eel temperature $T_N \sim 270$ K.\cite{mangpk:2004}

The RIXS data were collected in vertical scattering geometry with the X-ray spectrometer on beamline 9-ID-B, at the Advanced Photon Source, and using the 2~m arm configuration. The energy resolution (FWHM) was $0.32$~eV (Sec. \ref{ssc:nco:psi}) and $0.25$~eV (Sec. \ref{ssc:nco:scatgeom}).

The `elastic tails' due to the elastic peak and non-zero energy resolution are subtracted from the inelastic spectra above 1.5 eV energy loss. This is accomplished by (i) using the elastic peak to establish zero energy transfer, (ii) fitting energy-gain data (typically up to 3 eV) to the heuristic modified Lorentzian form ~ $1/(1+|w|^\alpha)$ where $w$ is energy-loss (with typical values of $\alpha$ in the range 1-2), (iii) and then subtracting the result of the fit from the energy-loss part of the spectrum.

\subsection{Azimuthal ($\psi$-) scans\label{ssc:nco:psi}}

\begin{figure}
\begin{center}
\includegraphics[width=3.5in]{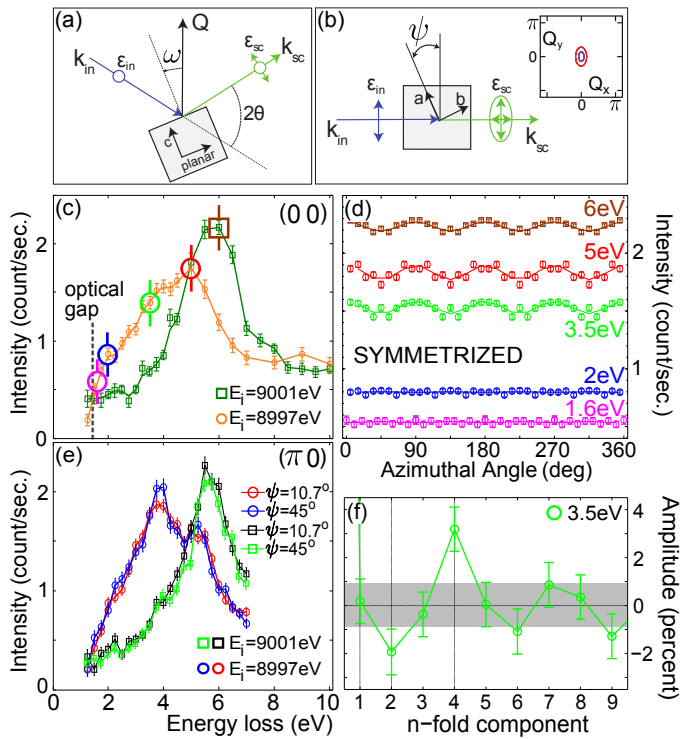}
\caption{(a) Side and (b) top view of the scattering geometry used throughout this paper. The polarization of the incident beam ($\sigma$) and the two possible polarization conditions for the scattered beam ($\sigma$ and $\pi$) are emphasized in both panels. Note that $\psi$-scans are collected at $\omega=0$ only (b) inset: two-dimensional (2D) Brillouin zone: the red line shows the full region of integration arising from the momentum resolution whereas the blue line shows its half width.
(c) Energy scans taken at the 2D zone center with incident energy $E_\textrm{i}$=8997~eV (circles) and 9001~eV (squares). The superposed colored circles and square show where the $\psi$-scans of corresponding color and energy are measured in (d) are measured. The charge-transfer gap measured with optical conductivity is indicated for comparison.\cite{onosey:2004}
(d) Azimuthal scans taken in a full circle around the 2D zone center at $1.625$, $2$, $3.5$, $5$~eV energy-loss (for $E_\textrm{i}=8997$~eV) and at $6$~eV energy-loss for ($E_\textrm{i}$=$9001$~eV) for $\vec{Q}=(0~0~9.1)$ and symmetrized with respect to the $90^{\circ}$ rotations and mirror planes of the underlying tetragonal structure.
(e) Energy scans with elastic tail subtracted (see text) taken at $\vec{Q}=(0~0.5~11.1)$ for different values of the azimuthal angle ($\psi$). The scans indicated by circles were taken at $E_\textrm{i}=8997$~eV whereas those indicated by squares were taken at $E_\textrm{i}=9001$~eV. Energy scans with $\psi=45^\circ$ are corrected for changes in self-absorption ($+5.2$\%) and incident polarization ($+0.8$\%) compared to $\psi=10.7^\circ$.
(f) Fourier components of the $\psi$-scan at $3.5$~eV energy-loss with amplitude given in percent of 0-fold (DC) component (not shown). The 4-fold component is well above the statistical noise level (gray region). There is also a 2-fold component present in the raw data.}
\label{fig:psiscan}
\end{center}
\end{figure}

In order to study the effects of photon polarization independently of the momentum transfer, we perform azimuthal scans, which rotate the incident-photon polarization within the CuO$_2$ planes, while keeping the energy- and momentum-transfer constant (the rotation axis is parallel to \textbf{Q}.\cite{murakamiy:1998}) This scattering geometry is illustrated in Fig. \ref{fig:psiscan}a, with $\omega=0$, and in Fig. \ref{fig:psiscan}b. The azimuthal angle $\psi$ is the scanned variable.

Figure \ref{fig:psiscan}c presents characteristic RIXS spectra taken at two different incident energies. These line scans show electronic excitations in the 1-10~eV energy-loss range associated with the electronic structure of the strongly-correlated CuO$_2$ plane, and they indicate the energies where $\psi$-scans are made. At the incident energy $E_\textrm{i} = 8997$~eV, the chosen values span from $1.625$~eV, just above the optical charge-transfer gap,\cite{onosey:2004} to the maximum of the inelastic intensity at 5~eV. At $E_\textrm{i} = 9001$~eV, we study the molecular orbital excitation~\cite{kimyj:2004c} by measuring at 6~eV.
The raw azimuthal scans are symmetrized to be consistent with the underlying tetragonal symmetry of the CuO$_2$ planes. This symmetrization consists of folding back the data within the 0-45$^\circ$ arc and then averaging it, the results are shown in Fig. \ref{fig:psiscan}d.
This procedure averages out all non-4-fold components exactly which, in practical terms, filters out experimental noise and leaves only the intrinsic electronic properties. For $E_\textrm{i}=8997$~eV, the resulting amplitudes of the 4-fold oscillations are $-0.8\pm3.5\%$ (at 1.625~eV),  $-0.2\pm0.9\%$ (2~eV), $3.2\pm0.9\%$ (3.5~eV), and $3.2\pm0.9\%$ (5~eV), while for $E_\textrm{i}=9001$~eV, we find $1.6\pm0.5\%$ (6~eV). Only the last three 4-fold oscillations, albeit small, are statistically significant (better than $3\sigma$) and inconsistent with being statistical noise. Fourier analysis of the $\psi$-scan at 3.5~eV (E$_i=8997$~eV) is presented in Fig. \ref{fig:psiscan}f, where the amplitude is in percent of the 0-fold (DC) component. Of the four physical components (0-,1-,2-,4-fold), only the 0-,~2-,~and~4-fold components are above the statistical noise level.\cite{Note1} The 2-fold component is systematically observed in all~$\psi$-scans, but filtered out by the symmetrization procedure.

We can rule out the extrinsic effects of the experimental configuration details as the origin of the observed 4-fold oscillations. Extrinsic effects due to sample self-absorption and the X-ray beam `footprint' are expected to be 1- and 2-fold respectively. In order to minimize these effects, we aligned the scattering surface normal along the $\psi$-scan rotation axis (i.e., $\omega=0$ in Fig. \ref{fig:psiscan}a) and we made sure that the X-ray beam spot was contained by the sample surface. After this procedure, no measurable 1-fold component was observed, but a systematic 2-fold component of $2\%$ remained. This 2-fold component is observed both within our RIXS signal and with a fluorescence monitor placed below the analyzer. Its strong dependence on the X-ray beam position within the scattering surface confirms that it is a footprint effect. Furthermore, extrinsic effects should be independent of both the selected intermediate state and the final electronic state that is probed and, as a result, be equally prominent for all energy-loss and incident-photon energies. This is inconsistent with the variability of the 4-fold oscillation amplitude we observe. The rotation of the anisotropic momentum-resolution-ellipsoid in reciprocal space (see Fig.\ref{fig:psiscan}b), might create an artificial 4-fold oscillation pattern in the $\psi$-scans, but a calculation based on the ellipsoid's shape and on the anisotropy of the momentum dependent inelastic signal gives an upper bound on this effect to less than $0.05\%$, much smaller than the observed oscillation amplitudes.

We can furthermore rule out that the 4-fold oscillations result from the resonant nature of RIXS, i.e., from the details of the resonantly excited intermediate state. In our experiment, the intermediate state consists of a $1s$ core hole and an in-plane $4p$ photoelectron created by the absorption of the incident photon. A local distortion of the lattice could introduce a 2-fold oscillation by splitting the $4p_x$ and $4p_y$ levels, as in the case of the manganites,\cite{murakamiy:1998} but the crystal structure of Nd$_2$CuO$_4$ is tetragonal and the $4p_x$ and $4p_y$ levels are degenerate. The $4p$ states mix non-locally with the $3d_{x^2-y^2}$ state, which introduces a 4-fold component to the intermediate state. On the other hand, two factors make this effect negligible: quadrupole transitions are typically two orders of magnitude weaker than dipole transitions and the $3d$-$4p$ mixing is weak because of small overlap integrals and a large energy separation between the two bands. In addition, an intermediate-state effect should be independent of energy-loss value and be seen equally in all measured spectra which is inconsistent with what we observe.

We conclude that the 4-fold oscillation is a property of the electronic excitations in the final state. That is, as we vary the energy-loss value, we probe varying admixtures of final states with different symmetries.

Since the incident-photon polarization is kept within the CuO$_2$ planes during the azimuthal rotations, the $4p$-as-spectator approximation predicts the RIXS signal to be independent of $\psi$. While this accurately describes approximately 95\% of the inelastic signal, the observed 4-fold oscillations with peak-to-peak amplitudes up to $6.4\%$ are evidence of excitations created by the interaction of the $4p$ photoelectron with the valence system. Because the $4p$ photoelectron can transfer angular momentum, these excitations can be of a different nature than those created by the $1s$ core-hole. In indirect $K$-edge RIXS, this result constitutes the first evidence of scattered photon polarization effects beyond the $4p$-as-spectator approximation.

In Fig. \ref{fig:psiscan}e, we extend our analysis beyond the zone center. This figure shows the $\psi$-angle dependence of the inelastic spectra at ($0$ $\pi$) at E$_i=8997$~eV and $9001$~eV. The molecular orbital excitation at $5.6$~eV measured for E$_i=9001$~eV has a weak $\psi$ angle dependence: it is approximately $6$\% weaker for $\psi=45^\circ$ than for $\psi=10.7^\circ$. There might also be a $\psi$-dependent feature for E$_i=8997$~eV at approximately $3.8$~eV.

\subsection{Scattering-geometry dependence\label{ssc:nco:scatgeom}}

Scattering-geometry dependence of the RIXS signal can have three different origins: sample self-absorption, incident and scattered photon polarization effects, and momentum-transfer effects. Since all three generally affect the signal simultaneously, it is difficult to separate their contributions. In Sec. \ref{ssc:nco:psi}, we used a particular scattering geometry that allows for the polarization degrees of freedom to be varied independently of the momentum transfer while minimizing self-absorption effects. Alternatively, the scattering geometry dependence can be investigated by measuring the Brillouin zone dependence of the inelastic signal.

Recently, \citet{kimyj:2007} studied the Brillouin-zone dependence of charge-transfer excitations at the Cu $K$-edge of the Mott insulator \lco. Based on remarkable agreement between inelastic spectra taken at high symmetry points of different Brillouin zones, they concluded that the RIXS signal is the same in all Brillouin zones. However, a closer look at the experimental spectra shows an interesting energy-loss-dependent difference of approximately 10\% between spectra measured in different Brillouin zones.

\begin{figure}
\begin{center}
\includegraphics[width=3.5in]{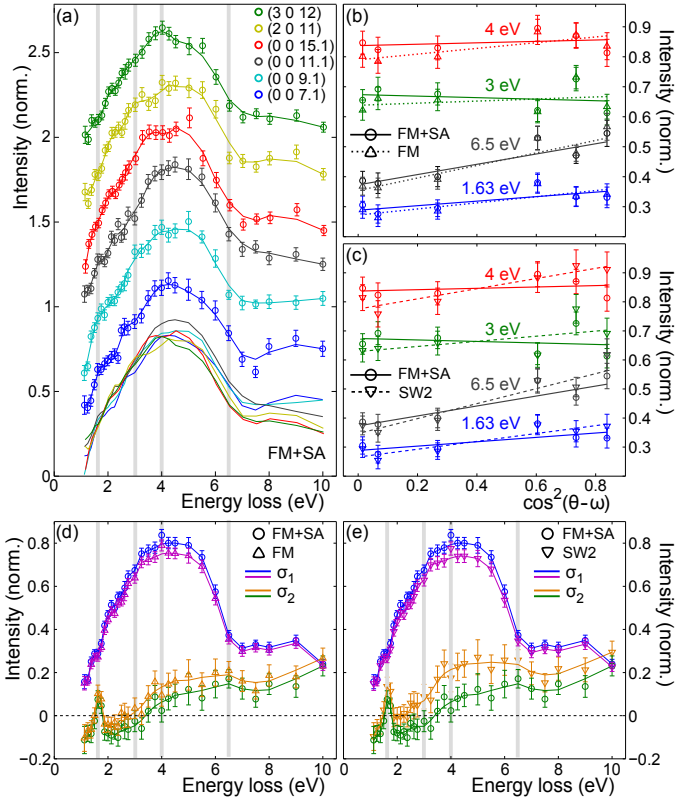}
\caption{(a) Two-dimensional zone-center spectra collected for a wide range of scattering geometries. Data sets are separated vertically by 0.3 units. The solid lines through the data are smooth interpolations and they are also overlaid at the bottom of the figure to highlight the evolution of the scattering geometry data. The data sets are scaled using the FM+SA method (see text for details). (b-c) Fits of the Brillouin-zone dependence to the linear form $y=\sigma_1+\sigma_2\cos^2(\theta-\omega)$ at 1.65, 3, 4, and 6.5 eV, comparing the FM+SA normalization procedure with (b) FM and (c) SW2 normalizations. (d-e) Fitted coefficients $\sigma_1$ and $\sigma_2$ as a function of energy-loss, comparing the FM+SA normalization procedure with (d) FM and (e) SW2 normalizations.}
\label{fig:augstack}
\end{center}
\end{figure}

In this Section, we study this subtle effect further by measuring the scattering-geometry dependence of the related Mott insulator Nd$_2$CuO$_4$. Inelastic line scans (energy-gain side subtracted) taken at the zone center of six different zones are shown in Fig. \ref{fig:augstack}a. The proper normalization of each spectrum is important as it allows the comparison of relative intensities even between widely different scattering geometries.

In order to separate extrinsic effects from intrinsic features, we compare four different normalization techniques. The first (FM) consists of using the fluorescence signal as a monitor. The second (FM+SA) adds the self-absorption correction described in Appendix \ref{sc:norm}. The third (SW1) and fourth (SW2) both use the fluorescence signal as a monitor and further normalize the spectra by the integrated inelastic spectral weight in a fixed energy-loss range, between 1.4-2.9 eV and 1.9-2.4 eV respectively. The resulting normalization factors are compiled in Table \ref{tab:norm} where the `raw' fluorescence monitor factors are separated from the different corrections using either self-absorption or integrated spectral weight. As expected, $(0~0~L)$ scattering geometries (where $\omega=0$) all have approximately the same normalization factor, except for $(0~0~7.1)$, where the X-ray footprint starts to be limited by the scattering surface size. The $(H~0~L)$ grazing-incidence normal-emission geometries (where $\omega<0$) suffer less self-absorption and have lower normalization factors accordingly. Based on the tabulated values, the fluorescence monitor provides the largest contribution to the normalization ($\approx20$\%), which validates its use as a first-order self-absorption correction. However, the additional self-absorption correction described in Appendix \ref{sc:norm} is not negligible ($\approx5$\%) and should be used.

\begin{table}
\setlength{\tabcolsep}{1.1mm}
\begin{tabular}{lccccccc}
              & & Raw & & \multicolumn{4}{c}{Correction factor}\\\cline{3-3}\cline{5-8}
\textbf{Q}    & & FM  & & FM & FM+SA & SW1   & SW2 \\
\hline
(0 0 7.1)     & & 1.185& & 1 & 0.960 & 1.081 & 1.091\\
(0 0 9.1)     & & 0.995& & 1 & 0.984 & 1.065 & 1.060\\
(0 0 11.1)    & & 1.000& & 1 & 1.000 & 1.000 & 1.000\\
(0 0 15.1)    & & 0.930& & 1 & 1.023 & 1.065 & 1.003\\
(2 0 11)      & & 0.966& & 1 & 1.043 & 1.072 & 0.967\\
(3 0 12)      & & 0.774& & 1 & 1.049 & 1.077 & 1.017\\
\hline
\end{tabular}
\caption{Normalization factors for six different scattering geometries. The raw fluorescence monitor (FM) normalization factors are compared to different correction methods (see text for details).}\label{tab:norm}
\end{table}

We analyze the scattering geometry dependence by dividing each of the six energy spectra into energy-loss bins, each bin half the size of the experimental resolution ($0.125$~eV). Within each bin, the scattering-geometry dependence of the intensity is fitted to a linear form ($y=\sigma_1+\sigma_2x$), where the dependent variable ($x$) is either $L$, $|Q|^2$, or $\cos^2(\theta-\omega)$ (see Fig. \ref{fig:psiscan}a for the definition of $\theta$ and $\omega$). Each fit is performed using the four different normalization procedures described above.

The quality of each linear fit is characterized by a reduced chi-square $\tilde{\chi}^2$ (the total chi-square divided by the number of degrees of freedom $d$). The chi-square values and degrees of freedom can be summed across the energy bins to compose a collective reduced chi-square $\tilde{\chi}^2=\sum_{i}\chi^2_i/\sum_id_i$ ($i$: bin index) which characterizes the fit functions' ability to represent the observed Brillouin-zone dependence across the entire data set. The goodness-of-fit indicator ($G$) is also calculated.\cite{Note2}

The values of these two indicators, for each combination of fit function and normalization procedure, are compiled in Table \ref{tab:chi2gi}. The $\cos^2(\theta-\omega)$ linear dependence provides the best fit to our data as it robustly yields the lowest $\chi^2$ and the highest $G$ values, independently of the normalization procedure. The $L$ and $|Q|^2$ fits are poorer, no matter what normalization procedure is used.

\begin{table}
\setlength{\tabcolsep}{1.1mm}
\begin{tabular}{lccc}
                          & $\cos^2(\theta-\omega)$ & $Q^2$ & $L$ \\\cline{2-4}
$\boldmath{\chi}^2$       \\\cline{1-1}
FM       & 1.433 & 2.032 & 2.033\\
FM+SA    & 1.513 & 1.747 & 1.731\\
SW1      & 1.179 & 1.558 & 1.543\\
SW2      & 1.138 & 2.062 & 2.030\\
\hline
\raisebox{-1ex}{\textbf{$G$} (\%)}\\\cline{1-1}
FM       & \phantom{1}0.13 & 1.9$\times$10$^{-8}$  & 1.8$\times$10$^{-8}$\\
FM+SA    & \phantom{1}0.02 & 7.7$\times$10$^{-5}$  & 1.2$\times$10$^{-4}$\\
SW1      & \phantom{1}8.81 & 8.8$\times$10$^{-3}$  & 1.3$\times$10$^{-2}$\\
SW2      & 14.31 & 7.2$\times$10$^{-9}$  & 2.0$\times$10$^{-8}$\\
\hline
\end{tabular}
\caption{Reduced chi-square ($\chi^2$) and goodness-of-fit ($G$) values for each combination of fit function and normalization procedure.}\label{tab:chi2gi}
\end{table}

Examples of the linear $\cos^2(\theta-\omega)$ fits are presented in Fig. \ref{fig:augstack}b-c and compare different normalization procedures. The energy-loss values chosen for this comparison span the spectral range of our data and correspond to the thick and gray vertical lines in Fig. \ref{fig:augstack}a,d-e. In Fig. \ref{fig:augstack}d-e, the resulting fit parameters (intercept $\sigma_1$ and slope $\sigma_2$) for each energy bin are compared for three of the four different normalization procedures.

The choice of normalization procedure affects the extracted amount of Brillouin zone dependence. For example, the FM+SA correction in Fig. \ref{fig:augstack}b,d adequately includes the effects of scattered-photon polarization: the self-absorption is larger for normal emission (in-plane scattered photon polarization, $\cos^2(\theta-\omega)\approx0$) than for grazing emission (mix of in-plane and out-of-plane photon polarization, $\cos^2(\theta-\omega)\approx1$). This difference is apparent form the polarization dependent X-ray absorption curves in Fig. \ref{fig:sa}c.

The SW2 normalization yields the lowest $\chi^2$ and largest $G$ which suggests that the 1.9-2.4~eV energy-loss region is Brillouin-zone independent. On the other hand, this normalization produces unphysically high values of $\sigma_2$ (Ref. \cite{Note3}) which suggests that the 1-3~eV energy-loss region in $\sigma_2$ is artificially reduced by a source of error beyond the self-absorption correction.

While a priori not unphysical,\cite{Note4} the negative offset in $\sigma_2$ below 3~eV (using FM+SA normalization) is probably an artifact the elastic-line subtraction. In our subtraction of the energy-gain side, we assume that the elastic line is symmetric. While this approximation is in principle valid, weak anisotropic elastic signal could `leak' into our RIXS spectra (due to the non-zero energy resolution) and introduce an artificial Brillouin-zone dependence of the signal at low energy loss. Errors in the fitted elastic line position can also introduce a weak Brillouin zone dependence. While this anisotropy is limited to low energy loss, an error in fitted position would be proportional to the RIXS spectrum's energy-loss slope and create artifacts at both high and low energy-loss. We emphasize that whereas the broad negative offset observed in $\sigma_2$ below 3~eV (within FM+SA) could be explained by a Brillouin-zone-varying asymmetry in the elastic line, neither of the above effects can create sharp features like the one observed at 1.65~eV.

Finally, a slight crystal misalignment could account for the overall larger intensity measured between 3-5~eV at (0~0~11.1), since the zone center spectrum at 8997~eV is a local minimum of inelastic intensity. On the other hand, the different spectral shape observed for (2~0~11) and (3~0~12) around 4~eV is not explicable by a crystal misalignment and could be evidence of a Brillouin-zone dependence not captured by the analysis presented in Fig. \ref{fig:augstack}.

While the many sources of error make the quantitative comparison of the spectra difficult, the quantities $\sigma_1$ and $\sigma_2$ exhibit robust features. The Brillouin-zone independent part of the RIXS spectrum ($\sigma_1$) consists of a broad feature centered at 4.5~eV, a shoulder at 2~eV (and possibly another at 1.4~eV), and a (linearly-extrapolated) onset of 0.8~eV. On the other hand, the Brillouin zone dependent part ($\sigma_2$) consists of a broad feature, centered around 5~eV with an onset around 2.5~eV, and its most interesting feature is a resolution-limited peak at 1.65~eV.

Because the spectra cannot be scaled to collapse onto one common curve, no matter what normalization is used, the Brillouin-zone dependent part ($\sigma_2$) cannot be spurious. The energy-loss dependence also rules out the resonant cross-section as the origin of the Brillouin-zone dependence, leaving only the properties of the measured electronic excitations to explain the effect.

For the employed scattering geometry, the $4p$-as-spectator approximation predicts no photon-polarization-based Brillouin-zone dependence of the inelastic signal. In contrast, the observed Brillouin-zone dependence is best fit by the $\cos^2(\theta-\omega)$ form which implies that the effect is photon-polarization-based and not momentum-based. As such, this observation implies that the $4p$ photoelectron interacts with the valence system during the RIXS process.

We note that the integration of the FM+SA normalized $\sigma_2$ spectral weight above 2.5~eV~\cite{Note5} sums to $15\pm2\%$ of the integrated $\sigma_1$ spectral weight. This departure from the $4p$-as-spectator approximation is larger than the $\psi$-scan peak-to-peak amplitude variations ($\sim\:6\%$), but it is of the same order of magnitude, which suggests a similar mechanism for both effects.

\section{Discussion\label{sc:disc}}

The analysis of the incident- and scattered-photon polarization dependence of the cross section is key to the study of the symmetry of electronic excitations in Raman scattering.\cite{devereauxtp:2007} Within (direct) soft RIXS, this has been used to distinguish the Zhang-Rice singlet (ZRS) excitation from local $d\rightarrow d$ excitations.\cite{okadak:2002, vanveenendaalm:2006} On the other hand, for (indirect) hard RIXS, it is unknown what excitation symmetries are measured and what their relative strengths are. It is furthermore unknown if RIXS obeys selection rules linking the incident- and scattered-photon polarizations and the underlying excitation symmetries.

While theoretical treatments have made assumptions about what types of excitations are measurable, they have not discussed selection rules explicitly. Treatments using joint-density-of-states-type cross sections limit the scattering from the core-hole to interband transitions between bands of the same point-group symmetry at $\vec{Q}\:=\:0$.\cite{takahashim:2009, markiewiczrs:2006} Furthermore, calculated RIXS spectra using one-band Hubbard models are automatically limited to states of $x^2-y^2$ local symmetry (the ZRS combination of oxygen orbitals is an $x^2-y^2$ combination of the O~$p_\sigma$ orbitals), so that the symmetry of the charge-transfer excitations is limited to $A_{1g}$ at $\vec{Q}\:=\:0$.\cite{tohyamat:2005b} On the other hand, non-$A_{1g}$ transitions have been suggested to explain new features in measured RIXS spectra: charge-transfers to non-bonding bands~\cite{lul:2005} and local $d\rightarrow d$ excitations~\cite{ellisds:2008} are both examples of such non-$A_{1g}$ transitions.

While it remains unclear if the polarization-based Raman selection rules can be applied to interpret indirect RIXS spectra at the Brillouin zone center (zero reduced $\vec{q}$), we test their applicability by comparing their predictions with our data. From the definition of the symmetry channels allowed by the tetragonal (D$_{4h}$) crystal structure of \nco (A$_{1g}$, A$_{2g}$, B$_{1g}$, B$_{2g}$, and E$_g$) and the evolution of the incident- and scattered-photon polarizations with the azimuthal angle ($\psi$) and the angular difference ($\theta-\omega$), we can write the photon-polarization-based Raman selection rules, at $\vec{Q}\:=\:0$, as a function of the scattering power within each allowed symmetry channel:
\begin{align}
I_{\text{Inel.}}\propto &\left[\sigma(A_{1g})+\sigma(A_{2g})+\sigma(B_{1g})+\sigma(B_{2g})\right]\nonumber\\
+&\left[\sigma(E_g)-\sigma(B_{1g})-\sigma(A_{2g})\right]\cos^2(\theta-\omega)\nonumber\\
+&\left[\sigma(B_{1g})-\sigma(B_{2g})\right]\cos^2(2\psi)\cos^2(\theta-\omega)\label{eqn:sym}
\end{align}
Since the scattered-photon polarization is not analyzed, we must include both the $\sigma$ and $\pi$ channels (polarization perpendicular and parallel to the scattering plane, respectively), as shown in Fig. \ref{fig:psiscan}a,b. Note that these rules inform us about the polarization dependence of these symmetry channels, but not about their relative size.

\begin{figure}
\begin{center}
\includegraphics[width=2.5in]{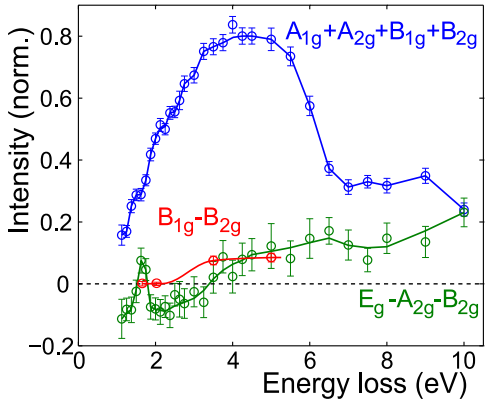}
\caption{Symmetry interpretation of the photon-polarization dependent and independent RIXS spectral weight for tetragonal Nd$_2$CuO$_4$. While composed of many different symmetry channels, the blue spectrum is most likely dominated by the $A_{1g}$ symmetry channel. See text for details.}
\label{fig:syms}
\end{center}
\end{figure}

At fixed $\theta$ and for $\omega\:=\:0$, Eq. \ref{eqn:sym} can be rewritten as $\sigma_0+\sigma_4\cos^2(2\psi)$ which is precisely the functional form of the 4-fold oscillations presented in Sec. \ref{ssc:nco:psi}. With a peak-to-peak 4-fold amplitude equal to $\left[\sigma(B_{1g})-\sigma(B_{2g})\right] \cos^2(\theta-\omega)$, these azimuthal oscillations can be interpreted in terms of the B$_{1g}$ and B$_{2g}$ symmetry channels. While we cannot determine the B$_{1g}$ and B$_{2g}$ amplitudes independently, our data suggest the presence of B$_{1g}$-type electronic excitations at $3.5$ and $5$~eV energy loss for $E_\textrm{i}=8997$~eV and at $6$~eV for $E_\textrm{i}=9001$~eV. Correcting the 4-fold peak-to-peak amplitude to account for the $\cos^2(\theta-\omega)$ dependence, we obtain an adjusted $B_{1g}-B_{2g}$ amplitude of $+8.7\%$ ($\sigma_4/\sigma_0$) between $3.5-5$eV for E$_i$=8997~eV. These adjusted data are shown in Fig. \ref{fig:syms}.

For $\psi=0$, Eq. \ref{eqn:sym} reduces to a cross section of the form $\sigma_1+\sigma_2\cos^2(\theta-\omega)$, where $\sigma_1=A_{1g}+A_{2g}+B_{1g}+B_{2g}$ and $\sigma_2=E_g-A_{2g}-B_{2g}$, which is precisely the function that best fits the Brillouin-zone dependence presented in Sec. \ref{ssc:nco:scatgeom}. This suggests that the decrease in spectral weight toward backscattering observed here for \nco and by \Citet{kimyj:2007} for \lco can be interpreted in terms of $E_g$ excitations at high energy-loss. These symmetry assignments for $\sigma_1$ and $\sigma_2$ are reproduced in Fig. \ref{fig:syms}.

In tetragonal symmetry, the incident- and scattered-photon polarizations of an $E_g$ excitation correspond to $4p$ crystal-field eigenstates with different energies (e.g., x$\rightarrow$z or z$\rightarrow$x), unlike the $A_{1g}$, $A_{2g}$, $B_{1g}$, $B_{2g}$ symmetry channels where the incident and outgoing $4p$ crystal-field eigenstates are degenerate (and planar).  This creates a unique resonance profile~\cite{doringg:2004, abbamontep:1999a, lul:2006, kimyj:2007} for $E_g$ excitations which an increase or decrease of intensity compared to the other symmetry channels.

For example, in the scattering geometry used here, at the incident energy $E_i=8997$~eV, $E_g$ and $B_{1g}$ excitations have the same incident-photon polarization resonance but their scattered-photon polarization resonances differ. For $B_{1g}$ excitations, the scattered-photon polarization is $\epsilon_f=\vec{x}$ and the final energy resonance is at $E_f=8997$~eV whereas for $E_{g}$ excitations, the scattered-photon polarization is $\epsilon_f=\vec{z}$ and there are scattered-photon resonances at $E_f=8985$, $8993$, and $8997$~eV according to the XAS data in Fig. \ref{fig:sa}. Around $4$~eV and $12$~eV energy-loss, the scattered photon is tuned to an out-of-plane resonance and $E_g$ excitations will be enhanced compared to $B_{1g}$ excitations. In this case, assuming that both symmetry channels have a comparable density of states, we estimate that $E_g$ excitations are comparatively enhanced by a factor of $2$-$5$. If we correct for this enhancement, the observed $15$\% Brillouin-zone dependence becomes a $3$-$8$\% effect, more comparable to the observed $8.7$\% azimuthal-angle dependence.

For the scattering geometry used here, the core-hole is expected to create the majority of the inelastic signal because it has the strongest effect on the valence system. We suggest that the Brillouin-zone-independent and $\psi$-angle independent spectral weight is created by core-hole scattering and consists of $A_{1g}$ symmetry. We note, however, that this contribution can be reduced to zero in certain scattering geometries because its cross-section follows the $4p$-as-spectator approximation discussed in Sec. \ref{sc:pol}.

While we have shown that the Raman selection rules can accurately model the observed photon-polarization effects and suggest symmetry assignments for different inelastic features, these assignments should be supported by a theoretical understanding of the excitations. The molecular orbital (MO) excitation at the poorly-screened state is understood to be of $A_{1g}$ symmetry, which is consistent with its weak $\psi$-angle dependence at $(0~\pi)$ (see Fig. \ref{fig:psiscan}e). While the MO excitation does show a weak 4-fold oscillation at the zone center, a comparison of the $(0~\pi)$ and $(0~0)$ spectra suggests that the 4-fold oscillation might instead be a property of a momentum-dependent shoulder that disappears away from $(0~0)$.

In principle, $d\rightarrow d$ excitations provide a testing ground for the validity of conventional Raman selection rules within RIXS, and such excitations have been well studied with soft X-ray RIXS at the Cu $L$-~\cite{ghiringhellig:2004, ghiringhellig:2009, haraday:2002, dudalc:2000b,dudalc:1998} and $M$-edges,\cite{kuiperp:1998} with optical absorption,\cite{perkinsjd:1993} with large-shift Raman scattering,\cite{salamond:1995} and with different theoretical methods.\cite{middlemissds:2008,degraafc:2000}

Although for Nd$_2$CuO$_4$ only the $A_{2g}$ ($d_{x^2-y^2}\rightarrow d_{xy}$) Cu crystal-field excitation has been observed around 1.4~eV,\cite{salamond:1995} the crystal-field excitations of tetragonal \scoc have been extensively studied~\cite{kuiperp:1998,perkinsjd:1993,ghiringhellig:2004,middlemissds:2008} and their energies should be similar to those of Nd$_2$CuO$_4$. For the latter, Raman selection rules suggest that the resolution-limited feature we observe at $1.65$~eV (Fig. \ref{fig:augstack}) is of $E_g$ symmetry. The $E_g$ crystal-field excitation ($d_{x^2-y^2}\rightarrow d_{xz}$) in \scoc has an energy of $1.7$~eV which, supports this symmetry assignment and the applicability of the Raman selection rules.

A complete determination of the $d\rightarrow d$ excitations in tetragonal \nco should be possible with the experimental methods described in this paper, i.e., without analyzing the scattered-photon polarization, since only three types of $d\rightarrow d$ excitations are possible (the $B_{2g}$ symmetry change does not exist within the Cu $3d$ orbitals). For example, with better energy resolution, the $A_{2g}$ excitation should be observable at $1.4$~eV with a Brillouin-zone dependence study, whereas the $B_{1g}$ ($d_{x^2-y^2}\rightarrow d_{3z^2-r^2}$) excitation should be observable with a $\psi$-angle dependence study.

Charge-transfers from the non-bonding oxygen bands to the upper Hubbard band, which have been used to explain the multiplet of inelastic features between 2 and 6~eV,\cite{lul:2005} are expected to be of non-$A_{1g}$ symmetry. As studied with ARPES,\cite{pothuizenjjm:1997, haynr:1999, damascellia:2003} O $2p$ non-bonding bands have approximately 1.5~eV more binding energy than the ZRS band. Accordingly, the RIXS charge transfer to such bands should start at an energy 1.5~eV above the overall onset of RIXS excitations. In Nd$_2$CuO$_4$, the onset of excitations is approximately $0.8$~eV. On the other hand, the 4-fold azimuthal-scan oscillations start between 2 and 3.5~eV, and the broad feature in the Brillouin-zone dependent part starts at around 2.5~eV. Both have onsets approximately 1.5~eV above 0.8~eV and could be interpreted as $B_{1g}$ and $E_g$ charge-transfers to oxygen non-bonding bands, respectively.

In their EELS study of Sr$_2$CuO$_2$Cl$_2$, \citet{moskvinas:2002} identify many charge-transfer excitations, three of which have A$_{1g}$ symmetry and should be RIXS active. The two strongest excitations are around $8$~eV, but do not have obvious RIXS equivalents. The third $A_{1g}$ excitation at $2$~eV, just above the optical gap, is identified as the Zhang-Rice singlet. This may correspond to the $2$~eV feature in RIXS, which is apparent as a weak shoulder in Fig. \ref{fig:augstack}d-e and is seen as a clear peak in \lco and Sr$_2$CuO$_2$Cl$_2$. This suggests that the 2 eV feature in RIXS is of $A_{1g}$ symmetry. An exact diagonalization calculation~\citet{tohyamat:2006} of charge-transfer excitations within a one-band Hubbard model of the CuO$_2$ plane agree quite well with our data and the $A_{1g}$ symmetry assignment of the 2 eV feature. The calculation shows a peak at $2$~eV that is exclusively of A$_{1g}$ symmetry, and a continuum of excitations between $2-3.5$~eV that is of predominantly B$_{1g}$ character, in good agreement with the symmetry interpretation of our data. The relative intensity of these two features disagrees with our observations, but quantitative agreement is not expected, since the resonant and non-resonant cross-sections are very different.

In order to better understand the symmetry selectivity of the RIXS cross section, it is worthwhile to compare it to well-established probes. On the other hand, symmetry-selectivity differs from probe to probe, which renders direct comparisons hazardous. For example, in crystals with an inversion center, optical conductivity measures a current-current correlation function only sensitive to \textit{ungerade} excitations, while electron-energy-loss-spectroscopy (EELS) and IXS measure a density-density correlation function that is sensitive to both \textit{ungerade} and \textit{gerade} excitations. Furthermore, the symmetry-sensitivity of these probes has a strong and non-periodic $\vec{Q}$-dependence~\cite{haverkortmw:2007, larsonbc:2007} in contrast to the nearly Brillouin-zone independent RIXS spectra.

\section{Conclusion}
By studying the photon-polarization dependence of indirect RIXS at the Cu $K$-edge of the tetragonal Mott insulator Nd$_2$CuO$_4$, we uncover anomalous excitations of two different types, which is evidence of scattered-photon polarization effects. While the majority (80\%) of the inelastic signal is describable by the $4p$-as-spectator approximation, the anomalous remainder constitutes the first evidence of RIXS excitations created by the interaction of the $4p$ photoelectron with the valence system.

The successful modeling of the observed azimuthal-scan 4-fold patterns and of the Brillouin-zone dependence by photon-polarization-based Raman selection rules suggests that these rules can be used to interpret zone-center RIXS spectra. Using these rules, we tentatively assign the sharp peak at $1.65$~eV in the Brillouin-zone-dependent spectral weight to an $E_g$ $d\rightarrow d$ excitation and the broad features above $2.5$~eV in both the Brillouin-zone-dependent and azimuthal-angle-dependent spectral weight to $E_g$ and $B_{1g}$ charge-transfers to non-bonding oxygen bands.

Establishing photon-polarization-based methods to characterize the electronic excitations' symmetry is a pivotal challenge for RIXS. Such methods should facilitate the interpretation of experimental spectra and help provide a better understanding of the underlying physics of the cuprates and other transition metal oxides. While a complete scattered-photon polarization analysis is currently prohibited by low count rates, ongoing instrumentation development and future increases in photon flux should soon make the full polarization analysis of the inelastic signal possible.

We would like to acknowledge valuable conversations with J. van den Brink, T.P. Devereaux, and K. Ishii. This work was supported by the DOE under Contract No. DE-AC02-76SF00515 and by the NSF under Grant No. DMR-0705086.

\appendix
\section{RIXS Normalization with a fluorescence monitor\label{sc:norm}}

Aside from effects intrinsic to the cross section, the RIXS scattering intensity is also modulated by the sample self-absorption, an extrinsic effect that depends on the scattering geometry.

In X-ray scattering, the absorption processes which determine the X-ray absorption length of a crystal are dominated by Auger emission of electrons and core-level fluorescence lines. Together, these processes' cross-sections dwarf the RIXS cross-section. As a result, when the X-ray beam travels through the sample, before and after the RIXS event, its intensity is attenuated. This attenuation has a strong dependence on the scattering geometry and determines the illuminated sample volume. The RIXS signal is proportional to the number of Cu atoms resonantly excited in this volume. This attenuation is referred to as self-absorption, and correcting for it is common practice for other spectroscopies scattering probes (for example, in EELS,\cite{finkj:1994} in EXAFS,\cite{boothch:2005} and in direct RIXS~\cite{dallerac:1997} for example) but has not yet become standard for indirect RIXS.

\begin{figure}
\begin{center}
\includegraphics[width=3.5in]{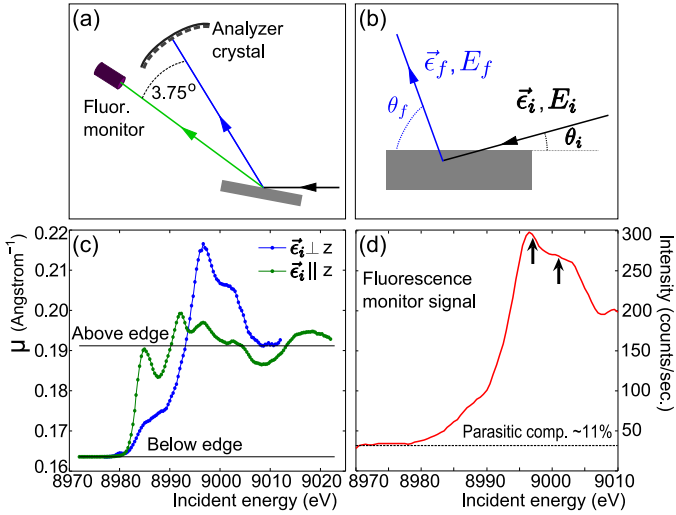}
\caption{(a) Position of the fluorescence monitor relative to the analyzer crystal (b) Self-absorption parameters in reflection geometry. (c) X-ray absorption spectroscopy (XAS) data (by partial fluorescence yield of the Cu $K_{\alpha1}$ emission line) in the Cu $K$-edge region. The black lines correspond to the calculated X-ray absorption coefficients above and below the Cu $K$-edge for Nd$_2$CuO$_4$. The measured XAS curves are scaled to follow the published curves by \citet{tranquadajm:1991}. (d) \nco absorption curve at the Cu $K$-edge measured with the fluorescence monitor as described in the text. The black arrows indicate incident energies of $8997$~eV and $9001$~eV.}
\label{fig:sa}
\end{center}
\end{figure}

To calculate this effect, we must know the X-ray absorption length and the scattering geometry. In reflection geometry, the RIXS intensity given by Eq. \ref{eqn:Reflec}. It includes the incident beam intensity $I_0$, the intrinsic RIXS scattering amplitude per Cu atom $F(E_i,\Delta E)$, the incident beam cross-sectional area $B_a$, the density of Cu atoms $\rho_{Cu}$, the absorption coefficient for the incident (scattered) photons $\mu_i$ ($\mu_f$) with polarization $\vec{\epsilon}_i$ ($\vec{\epsilon}_f$) and energy $E_i$ ($E_f$), as well as the scattering angles $\theta_i$ and $\theta_f$ given relative to the sample surface, as shown in Fig. \ref{fig:sa}b. The maximum scattering intensity is obtained for a photon with grazing incidence which is emitted perpendicular to the sample surface (an example is shown in Fig. \ref{fig:sa}b). In turn, the minimum intensity is observed for normal incidence and grazing emission angle. The reduction in RIXS signal from having a footprint (the beam spot on the sample) larger than the scattering surface is not included but is an important effect at grazing incidence.

\begin{align}
I&=\left(I_0\rho_{Cu}B_a\right)S\left(E_i,\vec{\epsilon}_i,E_f,\vec{\epsilon}_f\right)F\left(E_i,E_i-E_f\right)\nonumber\\
&=\frac{I_0\rho_{Cu}B_a}{\mu(\vec{\epsilon}_i,E_i)+\mu(\vec{\epsilon}_f,E_f)\frac{\sin(\theta_i)}{\sin(\theta_f)}}F\left(E_i,E_i-E_f\right)\label{eqn:Reflec}
\end{align}

In an ideal experiment, this formula could be used directly to normalize the RIXS spectra taken in different scattering geometries. In practice though, the footprint is highly dependent on the scattering surface and can affect the RIXS signal in non-trivial ways. To counteract this difficulty, we use a fluorescence detector put at a known position close to the analyzer crystal, as presented in Fig. \ref{fig:sa}a. After measuring the position of the sample surface relative to the crystal axes and tuning the fluorescence detector to an emission line (for example Cu $K_{\alpha1}$), we can calculate the scattering geometry dependence of both the fluorescence monitor (FM) signal ($S_{FM}$) and of the RIXS signal ($S_{RIXS}$). In order to correct for the self-absorption effect, which changes the intensity of the RIXS signal based on scattering geometry, we normalize with the following ratio:
\begin{equation}
I^C_{RIXS}=\frac{I_{RIXS}}{I_{FM}}\times\frac{S_{FM}}{S_{RIXS}}
\end{equation}
The measured fluorescence and RIXS intensities are $I_{FM}$ and $I_{RIXS}$. Simply dividing the RIXS signal by the FM signal is a first-order correction that can partially account for changes in the X-ray footprint on the sample. However, the exact correction must include variations in X-ray self-absorption based on the scattering geometry, the photon's polarization and energy all of which require calculating the self-absorption factors $S_{FM}$ and $S_{RIXS}$. Note that the $S_{FM}$ calculation includes the difference in location between the spectrometer's analyzer crystal and the fluorescence monitor, as shown in Fig. \ref{fig:sa}a. In practice, the further away the FM is placed from the analyzer crystal, the more difficult it will be to accurately account for positional differences in the calculation of the self-absorption correction. This can become a large source of systematic error in the normalization procedure, although this is not the case in the present experiment.

The fluorescence monitor signal described above is collected with a solid-state detector (Amptek). The signal is fed into a scaler and only photons within a fixed energy range are counted, in our case, in a 1~keV range around the strong Cu $K_{\alpha1}$ emission line at 8.1~keV. An absorption curve for \nco with in-plane incident-photon polarization collected with this fluorescence monitor is presented in Fig. \ref{fig:sa}d. The non-zero signal below the absorption edge is parasitic but is not the result of dark current within the detector. Instead, it is composed of two signals that leak into the energy integration window: elastic scattering  at 9~keV and neodymium $L$-edge fluorescence at 7~keV. Within our theoretical calculation, this parasitic component is tentatively compensated for by modeling it as 70\% elastic signal and 30\% neodymium fluorescence. Since the parasitic component does not vary congruently with the measured emission line, it constitutes a source of error and should in general be minimized.

In order to calculate the polarization dependent X-ray absorption coefficients ($\mu$) in the Cu K-edge region, we collected X-ray absorption data by partial fluorescence yield at the Cu $K_{\alpha1}$ emission line with an X-ray spectrometer.\cite{Note6} ($1.2$~eV resolution) The effect of self-absorption, typically important close to absorption edges, was calculated as described by \citet{carbonir:2005}, but was not significant here. In order to complete the construction of the absorption coefficients, the curves were then scaled to match the tabulated values for X-ray absorption in \nco above and below the Cu $K$-edge and following Ref. \citep{tranquadajm:1991}. The results are shown in Fig. \ref{fig:sa}c. While both the normalization procedure above the edge and the self-absorption correction of the fluorescence signal are sources of error, a conservative estimation of their effects is included in the normalization factor's error but, in our case, they are negligible.

This type of correction includes absorption effects due to changes of the scattered-photon energy that simply dividing by the fluorescence signal cannot account for. For example, based on Fig. \ref{fig:sa}c, an incident in-plane photon with an energy of $8997$~eV would see more absorption on its way into the sample than the scattered photon on its way out of the sample with $10$~eV energy-loss ($E_f=8987$~eV). Within an inelastic spectrum taken in these conditions and corrected by the fluorescence signal alone, the high-energy-loss response would be artificially larger than that at low energy, based solely on this change in the XAS coefficient.

Finally, the scattered photon can in principle have $\sigma$ or $\pi$ polarization which affects the calculated self-absorption correction. In vertical scattering geometry, at the in-plane $4p$ resonance, and within the bounds of the $4p$-as-spectator approximation, the scattered photon must be $\sigma$ polarized and we calculate the self-absorption correction based on this prescription. As seen in Sec. \ref{ssc:nco:scatgeom}, there is evidence that this approximation only describes part of the RIXS signal, with the complementary contribution in the $\pi$ polarization channel. This latter polarization channel, when isolated, can be normalized by modifying the calculated scattered photon X-ray absorption coefficient ($\mu_f$) accordingly and multiplying the $\sigma\rightarrow\pi$ scattering intensity by the ratio of the $\sigma\rightarrow\pi$ and $\sigma\rightarrow\sigma$ calculated self-absorption corrections:
\begin{equation}
f^{\sigma\rightarrow\pi}_{RIXS}=\frac{S^{\sigma\rightarrow\pi}}{S^{\sigma\rightarrow\sigma}}
\end{equation}
On the other hand, within the measured spectral range, the maximum correction is only $4\%$ of the Brillouin-zone \emph{dependent} signal and does not noticeably affect the normalization outlined above.

\end{document}